\documentclass[10pt,a4paper]{article}
\usepackage{graphicx,amsmath,amssymb,fixmath,theorem}

\usepackage{epsfig}

\usepackage{color}

\theoremstyle{break}
\theoremstyle{break}
\theoremstyle{break}

\title{Flexible and robust patterning by centralized gene networks.}

 \author{S.~Vakulenko$^1$ and O.~Radulescu$^2$ \\ \\
 \small  $^3$ Saint Petersburg State University of Technology and Design, St.Petersburg, Russia, \\
 \small  $^4$ DIMNP UMR CNRS 5235, University of Montpellier 2, Montpellier, France.}

\begin{document}

\maketitle

\centerline{\bf Abstract}

We investigate the possibility of programming arbitrarily complex space-time patterns,
and transitions between such patterns, by gene networks. We consider networks with two
types of nodes. The $v$-nodes, called centers,
are hyperconnected and interact one to another via $u$-nodes, called satellites.
This centralized architecture realizes a bow-tie scheme and possesses interesting properties.
Namely, this organization creates feedback loops that are capable to generate  any prescribed
patterning dynamics, chaotic or periodic, or stabilize a number of prescribed equilibrium states.
We show that activation or silencing of a node can sharply switch the network dynamics,
even if the activated or silenced node is weakly connected. Centralized networks can keep their flexibility, and still be protected against environmental noises.
Finding an optimized network that is both robust and flexible is a computationally hard problem in general,
but it becomes feasible when the number of satellites is large.
In theoretical biology, this class of models can be used to implement the
Driesch-Wolpert program, allowing to go from morphogen gradients to
multicellular organisms.

\section{Introduction}

The richness of Alan Turing's ideas hides 
somehow their unity. Is there a relation between
the ``chemical theory of morphogenesis'' (Turing 1952) and the ``universal machine'', or other, less known works,
such as
``Intelligent machinery'' (Turing 1969, Teuscher and Sanchez 2001) in which he anticipates random binary networks?
As emphasized by M.H.A. Newman (Newman 1955),
a common denominator of Turing's scientific work is the quest for a mechanical explanation of nature.
However, an even deeper unifying idea concerns the computability of nature and reciprocally,
how nature computes. Turing looked for a mechanical support for natural pattern computation and found
an analog machine, working by the chemical morphogens. Had he had known about gene networks, he would have probably
analyzed the computational capacity of these networks to make patterns and multicellular organisms.
In this paper we discuss a particular class of gene networks, the centralized or bow tie networks,
and show that they can ``compute'' multicellular organisms and comply with important desiderata of life such as
flexibility and robustness.

Flexibility and robustness are important properties of living systems in general, most particularly
observed during development from egg or embryo to a large, fully organized organism. Flexibility means the
capacity to change, when environmental conditions vary. Opposite to this, robustness is the capacity
to support homeostasis in spite of external changes. Intriguingly, biological
systems are in the same time robust and flexible.
Organization  of the body plan in development,
should be robust under unavoidable fluctuations of maternal gradients, embryo size,
and environment conditions (for instance, temperature). This is a viability condition.
On the other hand, developmental systems should be flexible in order to
produce a number of different patterns and complicated dynamics.

Pattern formation processes are important for
the body plan establishment via cell fate decisions in
multicellular organisms. Mathematical modeling of these phenomena
(Turing 1952, Meinhardt 1982, Murray 1993, Wolpert et al. 2002, Mjolness et al. 1991,
Reinitz and Sharp 1995, Page et al. 2005) is based on systems of reaction-diffusion equations,
sometimes with spatial inhomogeneous reaction terms.
Two different approaches, both considering
that laws of physics and chemistry are sufficient to account for making of
a multicellular organism, are fundamental in modeling of
body plan organization. The first approach, pioneered by the seminal
work of Turing (1952),
uses diffusion-driven instability as a patterning mechanism.
For Turing's mechanism, a spatial dependence of the reaction term is not necessary,
and the translation symmetry breaking needed for body plan organization results from
the Turing instability.
The second approach is based on Driesch-Wolpert positional information (Wolpert et al. 2002,
Wolpert 1970). The corresponding models can be also based on
reaction-diffusion equations, but in this case the models have
space dependent reaction terms and no translation symmetry,
therefore Turing instability is not needed. The main example of this type of
models is the gene circuit model (Mjolness et al 1991, Reinitz and Sharp 1995).
In this case, spatial organization is triggered by pre-patterns
of signaling molecules, generically called morphogen gradients.
It is remarkable that germs of this second approach can be found in
the conclusion of Turing's 1952 paper, where it is suggested that
``most of a organism, most of the time, is developing from one pattern
into another, rather than from homogeneity into a pattern''.

The body plan, considered as well defined, stable sequence of transitions
from one  pattern to another, can be encoded in a system of gene-gene
interactions or gene network. For instance, in the segmentation along the
anterior-posterior axis of Drosophila (fruit-fly) embryos, the
chemical support of the pre-pattern is the maternal bicoid gradient developed
in eggs soon after fecundation. This gradient induces spatially localized expression
of segmentation genes  (hunchback, kr\"uppel, giant, knirps, tailless, fushi tarazu, even skipped,
runt, hairy, odd skipped, paired, sloppy paired, etc.) forming
a gene network.
This gene network employs several types of interactions to stabilize the segmentation pattern.
Some of these interactions originate in {\em trans}, i.e., far, on the DNA sequence,
from the gene, and are due to transcription factors (TF)
and microRNAs (miRNA). Other interactions originate in {\em cis},
i.e., close, on the DNA sequence, to the gene. Indeed, the zygotic genome contains
cis-regulatory elements (CREM) controlling expression of the segmentation genes.
The gene circuit model (Mjolness et al 1991, Reinitz and Sharp 1995) accounts for part
of the {\em trans} interactions,
considering that segmentation genes are mutually regulated transcription factors.
The miRNA and the CREMs interactions are not represented in this model.
Although the role in stabilizing the development has been experimentally
proven for miRNAs(Li et al. 2009) and for CREMs (Ludwig et al. 2011), the mechanistic
details of these interactions are still unknown.
MiRNAs and CREMs can be abstractly considered as intermediate
nodes in a gene network, mediating interactions between transcription factors.
In such networks, TFs can be target hubs, being controlled by many CREMs, and also
source hubs, because they can bind to many CREMs. Similarly,
a few bioinformatics studies (Shalgi et al. 2007) suggest the
existence of many genes submitted to extensive miRNA regulation with
many TF among these target hubs. Direct testing of these interactions
have recently shown that important transcription factors
can be regulated by multiple miRNA (Tu and Bassler 2006, Martinez et al 2008,
Wu et al 2010, Peter 2010). Without excluding other applications of our mathematical
framework, we consider the TF-miRNAs networks as well as the TF-CREMs networks
as possible examples of centralized networks, or bow-tie networks.



The main goal of this paper is to show, by rigorous mathematical methods,
the following new results:

{\bf i}  centralized networks can create ``a multicellular organism'' consisting
of many specialized cells where the network dynamics within each cell can have a
different attractor,

{\bf ii} this pattern is robust under variations of morphogenetic fields; our system
performs trade-offs between flexibility and robustness,

{\bf iii} bifurcations between attractors can be obtained by gene silencing or reactivation.

These results, however, would be useless, without  algorithms that can resolve, in polynomial time $Poly(N)$,
(where $N$ is the gene number), the problem of a prescribed complicated and robust pattern construction (``computation of a
robust organism'', CRO problem). It is one of the key questions
in development, why evolution had a sufficient time to construct complicated organs and organisms (Darwin, Origins of Species, Chapter 6).
This problem is, in fact, a hard combinatorial one. Using new ideas in such problems,
  we show, under some assumptions, that

{\bf iv} for centralized networks with large hub connectivity, the CRO problem is feasible in polynomial  $Poly(N)$ time.

Similar ideas, that bow-tie connectivity can play a role in flexibility and robustness,
have been proposed by (Csete and Doyle 2004, Ma et al. 2007) in the context of metabolism,
but lacked mathematical proofs. In theoretical computer science, it was shown  that artificial
neural networks can simulate any Turing machine (Siegelmann and Sontag 1991, 1995).
Also, it was shown that networks can simulate any time trajectories
(Funahashi and Nakamura 1993) and any attractors (Vakulenko 1994, 2000, Vakulenko and Gordon 1998).

We extend these results to simulation of any spatio-temporal structure, with any
attractors. Since the pioneering ideas of Delbr\"uck (1949), it became well accepted that
differentiation and specialization of initially undifferentiated clone cells can be
understood via multiple dynamical structures and attractors (Thomas 1998). In particular, differences between  gene expression programs can be understood as differences
between attractors of dynamical gene networks.
At least mathematically, the possibility to control any
spatio-temporal pattern is equivalent to the possibility to organize any
multicellular organism.
Thus, we show that centralized networks can be used to
implement  Driesch-Wolpert positional information paradigm in order
to organize a multicellular organism. This organism consists of a number
of specialized cells, each cell type being dynamically characterized by distinct attractors.
The complexity of the attractors, that can be arbitrarily large, can be programmed
by gradients of morphogens.
Transitions between attractors can be performed by acting on key nodes of the network.
Contrary to previous theories of random networks (Kauffman 1969, Aldana 2003, Aldana and Cluzel 2003),
these key nodes do not have to be hubs. Furthermore, we show that patterning in such networks
is {\em maximally flexible} in the sense that it can produce any structurally
stable attractor. We also prove that (and show how) optimally flexible and robust structures
can be computed in polynomial time and can thus be easily attained by evolution.

The paper is organized as follows. Centralized networks are introduced in Section 2.
A first theorem (Proposition 2.3) concerns with
the flexibility of general centralized networks.
We show that these networks are capable to generate practically all structurally
stable prescribed dynamics, chaotic or periodic,
and can have
any number of equilibrium states.
Another key result (Theorem 2.5) can be interpreted, in biological terms,
as follows. The centralized networks are capable
to create a ``multicellular organism'', where each cell have a
prescribed time dynamics. This assertion can be considered as a mathematical realization of the Wolpert
approach since this intrinsic dynamics in a cell is predetermined only by the morphogen concentration in this cell.
In Section 3 we show that gene activation or silencing
can produce a sharp change of dynamics even if this gene is weakly connected
in the network (it is well known that a mutation in a hub can sharply change the dynamics, see Aldana 2003).
We show that in such a way one can obtain arbitrary bifurcations.
In Section 4 we consider the robustness of centralized networks and show how these can
acquire protection against environmental perturbations. We show that the design of a network
that is both flexible and robust can be stated as an optimization problem for a discrete
spin hamiltonian. When the number of satellites $N$ is large, the optimization problem
can be solved in polynomial time, $Poly(N)$.

\section{Centralized networks}

By centralized networks we mean networks that contain a few strongly connected nodes (hubs)
and a number of less connected, satellite nodes. A typical example is given by
scale-free networks (Albert and Barabasi  2002, Lesne 2006), that occur in many areas, in economics,
biology and sociology.
In the scale-free networks the probability $P(k)$ that a node is connected with $k$ neighbors,
has the asymptotics $Ck^{-\gamma}$, with $\gamma \in (2,3)$. Such  networks
typically contain a few hubs and a large number of satellite nodes.
Hence, scale-free networks are, in a sense, centralized.

In order to model dynamics of centralized networks we adapt a gene circuit model
 proposed to describe early stages of Drosophila (fruit-fly) morphogenesis
(Mjolness et al. 1991, Reinitz and Sharp 1995). To take into account the two types of the nodes,
 we use distinct variables $v_j$, $u_i$ for the centers and the satellites.
The real matrix entry
$A_{i j}$ defines the intensity of the action of
a center node $j$
on a satellite node $i$. This action can be either a repression $A_{i j} < 0$
or an activation $A_{i j} > 0$. Similarly, the matrices ${\bf B}$ and ${\bf C}$
define the action  of the centers on the satellites and the satellites on the centers, respectively.
Let us assume
that a satellite does not act directly on another satellite.
We also assume that
satellites respond more rapidly to perturbations and are more diffusive/mobile
than the centers.

Let $M, N$ be positive integers, and let ${\bf A}, {\bf B}$ and ${\bf C}$ be matrices of the sizes $N \times M, M \times M $
and $M \times N$ respectively. We denote by ${\bf A}_i, {\bf B}_j$ and ${\bf C}_j$ the rows of these matrices.
To simplify formulas, we use the notation
$$
   \sum_{j=1}^M A_{ij} v_j = {\bf A}_i v, \quad \sum_{l=1}^M B_{jl} v_l= {\bf B}_j v, \quad
   \sum_{k=1}^N C_{jk} u_k={\bf C}_j u.
$$

Then, the {\em gene circuit model} reads:
\begin{equation}
\frac{\partial u_i}{\partial t} = \tilde d_i \Delta u_i +
\tilde r_i \sigma\left( {\bf A}_i v + \tilde b_i m(x) - \tilde h_i\right) - \tilde \lambda_i u_i,
\label{cn1}
\end{equation}
\begin{equation}
\frac{\partial v_j}{\partial t} = d_j \Delta v_j +
r_j \sigma \left({ \bf B}_j v + {\bf C}_j u +
b_j m(x) - h_j\right) - \lambda_j v_j,
\label{cn2}
\end{equation}
where $m(x)$ represents the maternal morphogen gradient, $i=1,..., N, \ j=1,..., M$.
We assume that the diffusion coefficient $d_i, \tilde d_i$ and maximal production rates $r_i, \tilde r_i$ are non-negative:
$d_i, \tilde d_i, r_i, \tilde r_i \ge 0$.
Here the morphogenetic field $m(x)$ and unknown
gene concentrations $u_i(x,t), v_j(x,t)$ are defined in
a compact domain $x \in \Omega$ ($dim(\Omega) \leq 3$) having smooth boundary $\partial \Omega$,
$x \in \Omega$ and
 $\sigma$ is a monotone and smooth (at least twice differentiable) ``sigmoidal'' function such
that
\begin{equation}
\sigma(-\infty)=0, \quad \sigma(+\infty)=1.
\label {eq2.5}
\end{equation}
 Typical examples can be given
by
\begin{equation}
     \sigma(h)=\frac{1}{1 + \exp(- h)}, \quad \sigma(h)
     = \frac{1}{2} \left( \frac{h}{\sqrt{1+h^2}} + 1 \right).
\label {eq2.6}
\end{equation}
The function $\sigma(\beta x) $ becomes a step-like function as
its sharpness $\beta$ tends to $\infty$.

We also set the  Neumann boundary conditions
\begin{equation}
\nabla u_i(x,t) \cdot {\bf n}(x) =0, \quad
\nabla v_j(x,t) \cdot {\bf n}(x) =0, \quad (x \in \partial \Omega).
\label{Neumann}
\end{equation}
They mean that the flux of each reagent through the boundary is zero (here $\bf n$ denotes
the unit normal vector towards the boundary $\partial \Omega$ at the point $x$).
 Moreover, we set the initial conditions
 \begin{equation}
u_i(x,0)=\tilde \phi_i(x)\ge 0, \quad v_j(x,0)= \phi_j(x)\ge 0 \quad (x \in \Omega).
\label{initdat}
\end{equation}
 It is natural to assume that all concentrations are non-negative at the initial point,
 and it is easy to show that they stay non-negative for all times (see below).

Neglecting diffusion effects we obtain from (\ref{cn1}),(\ref{cn2}) the following {\em shorted system}:
\begin{equation}
\frac{\partial u_i}{\partial t} =
\tilde r_i \sigma\left({\bf A}_{i} v + \tilde b_i m(x) - \tilde h_i\right) - \tilde \lambda_i u_i,
\label{cn1s}
\end{equation}
\begin{equation}
\frac{\partial v_j}{\partial t} =
r_j \sigma\left({\bf B}_{j} v + {\bf C}_{j} u +
b_j m(x) - h_j\right) - \lambda_j v_j.
\label{cn2s}
\end{equation}
This is a Hopfield-like network model (Hopfield 1982)
with thresholds depending on $x$ (contrary to the Hopfield model,
the interaction matrices are not necessarily symmetric).
In this case we remove all boundary conditions (\ref{Neumann}).
If only $d_i=0$ we remove the corresponding boundary conditions for $v_i$.

\subsection{Existence of solutions}


Let us introduce some special functional spaces (Henry, 1981). Let us denote
$H=L_2(\Omega)^n$ the Hilbert space of the vector value functions $w$. This space is enabled by the standard $L_2$- norm defined by
$||w||^2= \int_{\Omega} |w(x)|^2 dx$, where $|w|^2=\sum w_i^2$. For $\alpha > 0$ we denote by
$H_{\alpha}$ the space consisting of all functions $w \in H$ such that the norm
$|| w||_{\alpha}$ is bounded, here
$
||w||^2_{\alpha}=|| (- \Delta + I)^{\alpha} w||^2.
$
These spaces have been well studied (see Henry 1981 and references therein).
The phase space of our system is ${\cal H}=\{w=(u, v) : u \in H, \ v \in H \}$, the corresponding
natural fractional spaces are denoted by $H_{\alpha}$ and ${\cal H}_{\alpha}$, here $H_{0}=H$ and ${\cal H}_0={\cal H}$.
Denote by $B_{\alpha}(R) $ the $n$-dimensional ball in $H_{\alpha}$ centered at the origin with
the radius $R$:
 $B_{\alpha}(R) =\{w: w \in H_{\alpha}, \ ||w||_{\alpha} < R\}$.

In our case all $f_i(w,x)$ are smooth in $w, x$, therefore, the standard technique (Henry 1981)
shows that solutions of (\ref{cn1}), (\ref{cn2})
exist locally in time and are unique. In fact, our system can be rewritten as an evolution
equation of the form
\begin{equation}
w_t= A w + f(w),
\label{eveq}
\end{equation}
where $f$ is a uniformly bounded $C^{1}$ map from
${\cal H}_{\alpha}$ to ${\cal H}$ (since  $\sup_{x \in \Omega}|w| \le c ||w||_{\alpha}, \, \alpha > 3/4$,
and the derivative $\sigma'(z)$ is   uniformly bounded  in $z$) and a linear self-adjoint negatively defined
operator $A$ generates a semigroup satisfying
the estimate $||\exp(At) w|| \le \exp(-\beta t)||w||$ with a $\beta > 0$.





Let us prove that the gene network dynamics is correctly defined for all $t$ and solutions are non-negative and bounded.
In fact, there exists an absorbing set ${\cal B}$ defined by
$$
{\cal B}=\{w=(u, v): 0 \le v_j \le r_j\lambda_j^{-1}, \ 0 \le u_i \le \tilde r_i \tilde \lambda_i^{-1}, \ j=1,...,M, \
 i=1, ..., N \}.
$$
 One can show, by super and subsolutions, that
\begin{equation}
\begin{split}
 0 \le u_i(x, t) \le \tilde \phi_i(x) \exp(-\tilde \lambda_i t) + \tilde r_i\tilde \lambda_i^{-1}
 (1-\exp(-\tilde \lambda_i t)), \\
 0  \le v_i(x, t) \le \phi_i(x) \exp(-\lambda_i t) + r_i\lambda_i^{-1}(1-\exp(-\lambda_i t)).
 \end{split}
\label{est0}
\end{equation}
Therefore, solutions of (\ref{cn1}), (\ref{cn2}) not only exist for all times $t$ but also they enter
 the set ${\cal B}$ at a time moment $t_0$ and then they stay in this set for all $t > t_0$. So, our system defines a global
dissipative semiflow (Henry, 1981).

\subsection{Reduced dynamics}

The key idea is to find a simpler asymptotic description of system dynamics. It is possible
 under some assumptions,
we suppose here that the $u$-variables are fast and the $v$-ones are slow.
We show then that the fast $u$ variables are captured, for large times,
by the slow $v$ modes. More precisely, one has
 $u= U(v) + \tilde u$, where $\tilde u$ is a small correction. This means
 that, for large times, the satellite dynamics is defined
almost completely by the center dynamics.

To realize this approach,
let us assume that the parameters of the system satisfy the following conditions:
\begin{equation}
 \ |A_{jl}|, |B_{il}|, |C_{ij}|, |\tilde h_i|, |h_j| < C_0,
\label{cn3}
\end{equation}
where $i=1,2,...,N, \ \ i,l=1,...,M,\ j=1,...,N$,
\begin{equation}
 0 < C_1 < \tilde \lambda_j, \quad \tilde d_j < C_2,
\label{cn4}
\end{equation}
\begin{equation}
 |b_j|, |\tilde b_i| < C_3, \quad \sup |m(x)| < C_4,
\label{cn31}
\end{equation}
and
\begin{equation}
 r_i=\kappa R_i, \quad \tilde r_i=\kappa \tilde R_i,
\label{cn41}
\end{equation}
where
\begin{equation}
 |R_i|, |\tilde  R_i|< C_5, \quad  \lambda_i=\kappa \bar \lambda_i, \ |\bar \lambda|  < C_6,
\label{cn41b}
\end{equation}
\begin{equation}
d_j=\kappa \bar d_j, \quad 0 < \bar d_j < C_7,
\label{cn37}
\end{equation}
where $\kappa$ is a small parameter, and where all positive constants $C_k$ are independent
of $\kappa$.

\vspace{0.2cm}

{\bf Proposition 2.1.}
{\em Assume the space dimension $Dim \Omega \le 3$.
Under assumptions (\ref{cn3}), (\ref{cn4}), (\ref{cn31}), (\ref{cn41})
 for sufficiently small $\kappa < \kappa_0$ solutions $(u, v)$ of (\ref{cn1}), (\ref{cn2}), (\ref{Neumann}),
  and (\ref{initdat})
  satisfy
\begin{equation}
u=U(x, v(x,t)) + \tilde u(x,t),
\label{cn5}
\end{equation}
where
the $j$-th component $U_j$ of $U$ is defined as a unique solution of the equation
\begin{equation}
\tilde d_j \Delta U_j - \tilde \lambda_j U_j= \kappa G_j(v),
\label{cn6}
\end{equation}
under the boundary conditions (\ref{Neumann}), where
$$
G_j=\tilde R_j
\sigma\left({\bf A}_{j} v(x, t) + \tilde
b_j m(x) - \tilde h_j \right)
$$
The function $\tilde u$ satisfies the estimates
\begin{equation}
||\tilde u|| + ||\nabla \tilde u|| < c_1\kappa^2 + R \exp(-
\beta t), \quad \beta > 0.
\label{cn8}
\end{equation}
The $v$ dynamics for large times $t > C_1 |\log \kappa |$ takes the form
\begin{equation}
\frac{\partial v_i}{\partial t} =\kappa F_i(u, v) + w_i,
\label{reddyn}
\end{equation}
 where $w_i$ satisfy
$$
||w_i|| < c_0\kappa^2
$$
and
$$
F_i(u, v) =\bar d_i \Delta v_i +
R_i \sigma\left(
{\bf B}_{i} v + {\bf C}_{i} U(x,v) + b_i m - h_i \right) -\bar \lambda_i v_i.
$$
Constants $c_0, c_1$ do not depend on $\kappa$ as $\kappa\to 0$ but they may depend on $R_i, \tilde R_i, C_i$.
}

A tedious proof of this assertion is basically straightforward; it is based on
well known results (Henry 1981) and is relegated to the Appendix.

An analogous assertion holds for shorted system (\ref{cn1s}),(\ref{cn2s}).
    In this case
the functions $U_i$ can be found by an explicit formula.
 Namely, one has
\begin{equation}
 U_i(x, v(x,t)) = \kappa V_i, \quad V_i=R_i \tilde \lambda_i^{-1}
\sigma\left({\bf A}_{j} v(x,t) + \tilde
b_j m(x) - \tilde h_j \right).
\label{cn10}
\end{equation}
For large times the reduced $v$ dynamics has the same form
(\ref{reddyn}) with $d_i=0$.

\subsection{Realization of prescribed dynamics by networks}

Our next goal is to show that dynamics (\ref{reddyn}) can realize, in a sense, arbitrary structurally
stable dynamics of the centers.
To precise this,
let us describe the method of realization of the vector fields
for dissipative systems (proposed by Pol\'a\v cik (1991),
for applications see, for example, (Dancer and Pol\'a\v cik 1999, Rybakowski 1994, Vakulenko 2000).
One can show that some
systems  possess
the following properties:
\vspace{0.2cm}

{\bf A} {\sl These systems generate global
semiflows $ S_{\cal P}^t$
in an ambient Hilbert or Banach phase
space $H$. These semiflows depend on some
parameters $\cal P$ (which could be elements of another
Banach space $\cal B$).
They have global attractors
and finite dimensional local attracting invariant
$C^1$ - manifolds $\cal M$
, at least for some $\cal P$}.

(Remark: in some cases, these manifolds can be even globally
attracting, i.e., inertial. Theory of invariant and inertial manifold
is well developed, see (Marion 1989, Mane 1977, Constantin et al 1989,
Chow and Lu 1988, Babin and Vishik 1988).

{\bf B} {\sl
Dynamics of
$S^t_{\cal P}$ reduced
on these
invariant manifolds is, in a sense, ``almost completely controllable''.
It can be described as follows. Assume the differential
equations
\begin{equation}
\label{2.1}
\frac{dp}{dt}=F(p), \quad F \in C^1(B^n)
\end{equation}
define a dynamical system in the unit ball
${ B}^n \subset {\bf R}^n$.

For any prescribed dynamics (\ref{2.1}) and any $\delta >0$,
we can choose suitable parameters ${\cal P}={\cal
 P}(n, F, \delta)$ such that

{\bf B1} The semiflow $ S_{\cal P}^t$ has a $C^1$-
smooth locally attracting invariant manifold ${\cal M}_{\cal P}$
diffeomorphic to ${B}^n$;

{\bf B2}
The reduced dynamics $ S_{\cal P}^t\vert_{{\cal M}_{\cal P}}$
is defined by equations
\begin{equation}
\frac{dp}{dt}=\tilde F(p, {\cal P}), \quad \tilde F \in C^1( B^n)
\label{tQ}
\end{equation}
where the estimate
\begin{equation}
|F -\tilde F|_{C^1({ B}^n)} < \delta
\label{est1}
\end{equation}
holds. In other words, one can say that, by $\cal P$,
the inertial dynamics can be specified to
within an arbitrarily small error.}

Thus, all robust dynamics (stable under small perturbations)
 can occur as inertial forms of these systems.  Such systems can be named {\em maximally
 dynamically flexible, or, for brevity, MDF systems}.

Such structurally stable dynamics can be ``chaotic''.
There is a rather wide variation in different definitions of
 ``chaos''. In principle, one can use here
any concept of chaos, provided that this is stable
under small $C^1$ -perturbations. To fix ideas, we shall use here,
following classical tradition
(Ruelle and Takens 1971, Newhouse, Ruelle and Takens 1971, Smale 1980, Anosov 1995), such a definition. We say that a finite
dimensional dynamics is chaotic if it
generates a
hyperbolic invariant set $\Gamma$, which is not a  periodic cycle or a rest point. For a definition of hyperbolic
sets see, for example, (Ruelle 1989); a famous example is given by the Smale horseshoe. If, moreover,
this set $\Gamma$ is attracting we say that $\Gamma$
is a chaotic (strange) attractor.
In this paper, we use only the following
basic property of hyperbolic sets,
  so-called Persistence (Ruelle 1989, Anosov 1995).
This means that the hyperbolic sets are, in a sense, stable(robust):
if (\ref{2.1}) generates the hyperbolic set $\Gamma$ and $\delta$
is sufficiently small, then dynamics (\ref{tQ}) also generates another
hyperbolic set $\tilde \Gamma$. Dynamics
(\ref{2.1}) and (\ref{tQ}) restricted to $\Gamma$
and $\tilde \Gamma$ respectively, are topologically orbitally equivalent
(on definition of this equivalence, see
Ruelle 1989, Anosov 1995).


   Thus, any kind of the chaotic hyperbolic sets
 can occur in the dynamics of the MDF systems, for example,
the Smale horseshoes, Anosov
flows, the Ruelle-Takens-Newhouse chaos, see (Newhouse, Ruelle and Takens 1971, Smale 1980, Ruelle 1989).
Examples of systems satisfying these properties can be given
 by some reaction diffusion
systems (Dancer and Pol\'a\v cik 1999, Rybakowski 1994, Vakulenko 2000).
Although not yet observed in gene networks,
structurally stable chaotic itineracy is thought to play a functional role
in neuroscience (Rabinovitch 1998).

 Let us apply this approach to network dynamics using the results of the previous section.
To this end, assume that (\ref{cn41}), \ref{cn41b}) and (\ref{cn37}) hold. Moreover, let us assume
\begin{equation}
 b_i=\kappa \bar b_i, \quad h_i=\kappa \bar h_i
\label{cbh1}
\end{equation}
\begin{equation}
 \lambda_i=\kappa^2 \bar \lambda_i, \quad d_i=\kappa^2 \bar d_i
\label{cbh2}
\end{equation}
where all coefficients $\bar b_i$ and $\bar h_i$
are uniform in $\kappa$ as $\kappa \to 0$.
These assumptions are useful for technical reasons.
We also assume that all
direct interactions between centers are absent, ${\bf B}={\bf 0}$.
This constraint is not essential but facilitates notation and calculations.

Since $U_j=O(\kappa)$ for small $\kappa$, we can use the Taylor expansion
for $\sigma$ in (\ref{reddyn}). Then these equations reduce to
\begin{equation}
\frac{\partial v_i(x, \tau)}{\partial \tau} =\bar d_i \Delta v_i
  + \rho_i ( {\bf C}_{i} V(x, v)
 + \bar b_i m(x) - \bar h_i) - \bar \lambda_i v_i +
\tilde w_i(x,t),
\label{cn11}
\end{equation}
where $\rho_i(x)=\bar r_i \sigma'(0)$, $i=1,2,...,M$
and $\tau$ is a slow rescaling time: $\tau=\kappa^2 t$.
 Due to conditions (\ref{cbh1}) and (\ref{cbh2}) corrections $\tilde w_i$ satisfy
$$
||\tilde w_i|| < c\kappa.
$$

Let us focus now our attention to non-perturbed equation (\ref{cn11}) with $\tilde w_i =0$.
Let us fix the number of centers $M$. The number of satellites $N$ will be considered as a parameter.

The next important lemma follows from known approximation
theorems of multilayered network theory, see, for example, (Barron 1993, Funahashi and Nakamura 1993).
\vspace{0.2cm}

{\bf {Lemma 2.2.}}
{ \em Given a number $\delta > 0$, an integer $M$ and a vector field $F=(F_1, ..., F_M)$ defined
on the ball $B^M=\{ |v| \le 1 \}$, $F_i \in C^1(B^M)$,
 there are a number $N$, a $N\times M$ matrix ${\bf A}$, a $M \times N$ matrix ${\bf C}$
and coefficients $h_i$, where $i=1,2,...,N$, such that
\begin{equation}
|F_j(\cdot) - {\bf C}_{j} W( \cdot)|_{C^1(B^M)} < \delta,
\label{cn14}
\end{equation}
where
\begin{equation}
 W_i(v) = \sigma\left({\bf A}_{i} v
  - h_i \right),
\label{cn40}
\end{equation}
where $v=(v_1, ..., v_M) \in  {\bf R}^M$.
}

This lemma gives us a tool to control network dynamics and patterns. First we
consider the case when the morphogens are absent. Formally, we can set
$\tilde b_i = \bar b_j = \bar d_i=0$. Assume $\bar h_i=0$. Then equations (\ref{cn11}) with $\tilde w_i=0$
reduce to
the Hopfield-like equations for variables $v_i \equiv v_i(\tau)$ that depend only on $\tau$:
 \begin{equation}
\frac{ dv_l}{d \tau} =
    {\bf K}_{l} W(v)
 - \bar \lambda_l v_l,
\label{hop}
 \end{equation}
where $l=1,..., M$, the matrix $\bf K$ is defined by
$K_{lj}=\rho_l C_{lj} R_j \tilde \lambda_j^{-1}$. The parameters $\cal P$ of (\ref{hop}) are $\bf K$, $M$, $h_j$ and $\bar \lambda_j$.

In this case one can formulate the following result.
\vspace{0.2cm}

{{\bf {Proposition 2.3.}}
{\em
Let us consider a $C^1$-smooth vector field $Q(p)$ defined on
a ball $B^M \subset {\bf R}^M$ and directed strictly inside this ball at the boundary $\partial B^M$:
\begin{equation}
F(p) \cdot p < 0, \quad p \in \partial B^M.
\label{cn15}
\end{equation}
Then, for each $\delta > 0$, there is a choice of parameters $\cal P$ such that (\ref{hop}) $\delta$ -realizes
the  system (\ref{2.1}). This means that (\ref{hop}) is a MDF system.
}

This proposition follows from the Prop. 2.1 and Lemma 2.2.

Prop. 2.3 implies the following important corollary:
 all structurally stable dynamics, including periodic and chaotic dynamics can be realized by centralized networks.
The  proof of this fact uses the classical results on the persistence of hyperbolic sets, and on the existence of
invariant manifolds
(Ruelle 1989), see
(Vakulenko 2000).


\subsection{Pattern and attractor control by Wolpert positional information}

Above we have considered a spatially homogeneous case.
Proposition 2.3 shows that a centralized network can approximate an arbitrary prescribed
 dynamics. Thus, it is shown that cells can be programmed to have arbitrarily
complex dynamics. By network rewiring or by interaction tuning,
one can switch between various types of dynamics. During development these switches
are position dependent, and induce cell differentiation into specific
spatial arrangements.

Let us show that the centralized networks, coupled to morphogen gradients,
can generate any spatio-temporal pattern as support for multicellular organization.
We consider shorted dynamics (\ref{cn1s}), (\ref{cn2s}) that is reasonable for cellularized developmental stages, where
cell walls prevent a free diffusion of regulatory molecules. Although other phenomena such
as cell signalling can also lead to cell coupling, we do not discuss these
effects here.

Assume cell positions are centered at the points $x\in {\cal X}=\{x_1, x_2,...,x_k\}$, $dim \Omega =1$,
${\cal X}$ is a discrete subset of $[0, L]$.
Let us show  that  eqs. (\ref{cn1s})-(\ref{cn2s})
can realize different dynamics at different points $x_l$
of the domain $\Omega=[0, L]$.

\vspace{0.2cm}
We can formulate now the following,

{\bf Theorem 2.5. (On translation of positional information into complex and variegated
cell dynamics, or programming of multicellular organism).
}
{\em Suppose $x \in [0, L] \subset {\bf R}$
and $m(x)$ is a strictly monotone smooth function.

Assume that $0 < x_1 < x_2 < ... < x_k < L$ and that
 $F^{(l)}(p), \ l=1,2,...,k$ is a family of $C^1$-smooth vector fields defined on a unit ball $B^M \subset {\bf R}^M$.
 We assume that each field defines a dynamical system, i.e.,
$F^{(l)}$ are directed inwards on the boundary $\partial B^M$.

Then, for each $\delta > 0$ there is a parameter $\cal P$ choice
such that for shorted dynamics (\ref{cn1s})-(\ref{cn2s})
one has
$$
  u=U(x, v) + \tilde u,
$$
 where
 $$
 |\tilde u| < C \exp(-\beta \tau) + c \kappa^2.
 $$
  For $x=x_l$ and for sufficiently large times the dynamics for $v(x_l, t)$ can be
  reduced to the form
\begin{equation}
\frac{dp_i}{d\tau}= \bar F_i(x_l, p),
\label{cn19}
\end{equation}
where
\begin{equation}
\sup_{p \in B^M} |\bar F(x_l, p) - F^{(l)}( p)| < \delta.
\label{cn19a}
\end{equation}
Here $p_i(\tau)$ can be expressed in a linear way via $v_i(x_l, \tau)$ by
$$
v_i(x_l, \tau)- \bar b m(x_l) = \rho_0 p_i(\tau).
$$
}
}

This theorem can be considered as a mathematical formalization of positional information
ideas. It extends Driesch-Wolpert theory by incorporating gene networks and coping
with their information processing role. Flexible gene networks have different dynamics
 and attractors, for different local concentrations of morphogens.
The attractor selection ensures the cell fate decision. Concerning the relation
between attractors and cell fate determination, see (Delbr\"uck 1949, Thomas 1998).

To prove this assertion, let us turn to eqs.
(\ref{cn11}), where, taking into account biological arguments
given above, we set $d_i = 0, \tilde b_i=0$. Let us set, to simplify formulas, $\rho_j=1, \bar h_j=0$ and $\bar \lambda_j=1$. Then
 $$
 V_j(v)=R_j \tilde \lambda_j^{-1}\sigma({\bf A}_{j} v - \tilde h_j).
 $$

Denote by $Q_i$ the sums
 $Q_i(v)=\sum_{j=1}^M C_{ij} V_j(v)= {\bf C}_i V$.
Removing the terms $\tilde w_i$ in (\ref{cn11}), one obtains that eqs. (\ref{cn11}) reduce to
\begin{equation}
\frac{\partial v_i(x, \tau)}{\partial \tau} =
  Q_i(v(x, \tau))
 + \bar b_i m(x) - v_i(x, \tau).
\label{cn20}
\end{equation}

Let us fix a $x=x_l \in {\cal X}$.
 Let us make the substitution $ v_i(x_l, \tau) = z_i(\tau) + \bar b m(x_l)$ in (\ref{cn20}) that gives
\begin{equation}
\frac{dz_i(\tau)}{d\tau} =
  Q_i(z + \bar b m(x_l))
    - z_i,
\label{cn21}
\end{equation}
where $\bar b=(\bar b_1, ..., \bar b_M)$, $i=1,..., M$.

Now we again use  approximation Lemma 2.2. Let us consider a family of vector fields
$C^1$-smooth vector field $F$ defined on
a unit ball $B^M=\{z \in {\bf R}^M, \ |z| \le 1\}$ and directed strictly inside this ball at the boundary $\partial B^M$:
\begin{equation}
F^{(l)}(z) \cdot z < 0, \quad z \in \partial B^M.
\label{inside}
\end{equation}

Assume  $m(x)$ is a strictly monotone function in $x$.
The main idea is as follows: since
all $m(x_l)=\mu_l$ and $m(x_j)=\mu_j$ are different for $j \ne l$,
the vector fields $Q^{(l)}(z)=Q(z + \bar b \mu_l)$
can approximate different vector fields $F^{(l)}(z)$ for  $l=1,..., k$
and for  $z$ such that $|z| < \rho_0$, where
$
\rho_0=\frac{1}{2}\min_{i,j, l, j \ne l} |\bar b_i| |\mu(x_j) - \mu(x_l)|$.

For each $\epsilon > 0$ we can find an approximation $Q$ satisfying
\begin{equation}
\sup_{|z| < \rho_0} |Q(z + \bar b m(x_l)) - (\rho_0 F^{(l)}\rho_0^{-1} z)+z)| < \rho_0 \epsilon,
\label{apr1}
\end{equation}
and
\begin{equation}
\sup_{|z| < \rho_0} |\nabla (Q(z + \bar b m(x_l))) - \nabla \rho_0 F^{(l)}(\rho_0^{-1} z)+z)| < \rho_0\epsilon.
\label{apr1d}
\end{equation}


Then equation (\ref{cn21}) reduces to
$$
\frac{dz}{d\tau}= \rho_0 F^{(l)}(\rho_0^{-1}z) + \rho_0 \epsilon \tilde F^{(l)}(z),
$$
where
$$
\sup_{z \in B^M} |\tilde F^{(l)}(z)| < 1, \quad \sup_{z \in B^M} |\nabla \tilde F^{(l)}(z)| < 1.
$$
We set $z_i= \rho_0 p_i$. This gives
\begin{equation}
\frac{dp}{d\tau}= F^{(l)}(p) + \epsilon \tilde F^{(l)}(p).
\label{Fl}
\end{equation}
Let us notice that, if $\epsilon$ is small enough, then for each index $l$,
due to assumption (\ref{inside}), the trajectory $p(t, p(0))$ stays in $B^M$ when the starting point lies in $B^M$: $p(0) \in B^M$.
Consequently, our approximations (\ref{apr1}) give vector fields that, by (\ref{Fl}), realize different dynamics
 for each $x_l$.

\section{Sharp genetic switch by satellite silencing/reactivation}

In the context of scale-free random networks,
it was proposed (Aldana 2003) that removing of a strong connected center can sharply change the network attractor.
Here we will show that one can obtain  transitions between all possible
structurally stable attractors by a single event acting on a specially chosen {\em weakly connected satellite}. Such a satellite
interacts only to one or two centers.
Such event may be, either deletion, silencing, or reactivation. Therefore, such a node can serve as a switch between two
kinds of network behavior.
Each of the type of behavior can  be defined, for example, by an attractor or several coexisting
attractors that can be fixed points, periodic or chaotic attractors.

To formalize these ideas  mathematically, let us consider a system of ordinary differential equations
\begin{equation}
   \frac{dp}{dt}=F(p, s), \quad p \in B^n \subset {\bf R}^n
\label{qs0}
\end{equation}
depending on a real  parameter $s$. Here $p=(p_1, p_2, ..., p_n)$, $B^n$ is the unit ball centered at $p=0$,
and $F$ is $C^1$-smooth vector field directed inside the ball at the ball boundary for each $s$
(see (\ref{inside})).   Let us
consider $s_0, s_1$ such that $s_0 \ne s_1$ and suppose that (\ref{qs0}) has different attractors
${\cal A}_0$ and ${\cal A}_1$ for $s=s_0, s=s_1$ respectively.

Consider, for simplicity,   the  gene circuit model (\ref{cn1}), (\ref{cn2}) without diffusion and space variables:
\begin{equation}
\frac{du_i}{dt} =
\tilde r_i \sigma\left( {\bf A}_i v  - \tilde h_i\right) - \tilde \lambda_i u_i,
\label{bb1}
\end{equation}
\begin{equation}
\frac{dv_j}{dt} =
r_j \sigma \left( {\bf C}_j u
 - h_j\right) - \lambda_j v_j,
\label{bb2}
\end{equation}
where $i=1,..., M+1$, $j=1,..., N$.
The parameters $\cal P$ of this system are $M, N, h_i, \tilde h_i$, $\lambda_i, r_i, \tilde r_j, \tilde \lambda_j$
and the matrices ${\bf A, C}$. We can assume, without loss of generality, that we eliminate (by silencing) the $M+1$-th  satellite node, $i=M+1$.
As a result of this elimination, we obtain a similar system with $i=1,..., M$ and shorted matrices ${\bf A}, {\bf C}$.
Of course, we can also consider the opposite event, which is to reactivate the $M+1$-th node and recover the initial system this way.

\vspace{0.2cm}

{\bf Theorem 3.1} {\em For each $\epsilon > 0$ there is a choice of the parameters $\cal P$
such that  system (\ref{bb1}), (\ref{bb2}) with $M$ satellite nodes $\epsilon$ -realizes
(\ref{qs0}) with $s=s_0$ and system (\ref{bb1}), (\ref{bb2}) with $M+1$ satellite nodes $\epsilon$ -realizes
(\ref{qs0}) with $s=s_1$.
}

To prove it, we use the following  extended system
\begin{equation}
   \frac{dp}{dt}=\rho F(p, s), \quad p \in B^n
\label{qs00}
\end{equation}
\begin{equation}
   \frac{ds}{dt}= f(s, \beta) - \nu s, \quad s \in {\bf R}
\label{qs01}
\end{equation}
 where $\nu > 1$ and $f(s)$ is a smooth function, $\beta, \rho >0$ are parameters.  Then equilibrium points $s_{eq}$ of (\ref{qs01}) are solutions of
  \begin{equation}
    f(s, \beta) = \nu s
\label{qss}
\end{equation}
The point $s_{eq}$ is a local attractor if $f'_s(s_{eq}) < \nu$. Let us denote   $s_{eq}(\beta_k)=s_k$, where $k=0,1$,  and let
 these roots of (\ref{qss}) be stable, i.e.,
$ f'(s_k) < \nu$.

Then, if $\rho > 0$ is small enough, and $s_k$ is a single stable rest point,
the fast variable $s$ approaches at $s_{eq}(\beta)$ and for large times $t$ the dynamics
of our system (\ref{qs00}), (\ref{qs01})  is defined by the reduced equations
\begin{equation}
   \frac{dp}{dt}=\rho F(p, s_{eq}(\beta)).
\label{qsr}
\end{equation}

Now let us set
 \begin{equation}
    f(s, \beta) = 2\beta^2 \sigma (b(s- h_0)), \quad h_0 < 0
\label{qf}
\end{equation}
where $b$ is a large parameter. Then $f$ is close to a step function with the step $2\beta^2$.
Therefore for $s_{eq}(\beta)$ one has the asymptotics $s_{eq}=2\beta^2\nu^{-1} + O(\exp(-b))$ as $b \to \infty$.
Thus, we can adjust parameters $\beta, b >0$ in such a way that
(\ref{qss}) has a single stable root $s_0$  and the equation
\begin{equation}
    f(s, \beta/\sqrt{2}) = \nu s
\label{qss2}
\end{equation}
also has a single root $s_1=\beta^2/\nu \ne s_0$.

Dynamics  (\ref{qs00}), (\ref{qs01}) with $f$ from (\ref{qf}) can be realized by a network (\ref{bb1}), (\ref{bb2})
in  such a way. We decompose all satellites $u_i$ into two subsets. The first set contains satellites
$u_1, u_2, ..., u_{M-1}$, the second one consists of the satellites $u_{M}, u_{M+1}$ ( to single out this
variables, let us denote $u_M=y_1, u_{M+1}=y_2$). The main idea
of this decomposition is as follows. We can linearize equations for the centers $v_j$
assuming that the matrix ${\bf C}$ is small and ${\bf B}=0$ (as above in Section 2).

The $y$ satellites realizes the dynamics (\ref{qs01}) by  a center $s$:
\begin{equation}
\frac{ds}{dt}=-\nu s +  \beta (y_1  +  y_2),
\label{ss}
\end{equation}
\begin{equation}
\frac{dy_k}{dt}=- y_k +  \beta  \sigma(b(s-h_0)),  \quad k=1,2.
\label{sy}
\end{equation}
Here we assume that $\nu, \beta$ is small enough, therefore, for large times this system reduces to
(\ref{qs01}) with $f$ defined by (\ref{qf}). We see that this dynamics bifurcates into
(\ref{qs01}) with $f = \beta^2 \sigma (b(s- h_0))$ if we remove $y_2$ in the right hand side of (\ref{ss}).

The rest of the equations, after a notation modification and linearization, take the following form
\begin{equation}
\frac{du_i}{dt} =
\tilde r_i \sigma\left( {\bf A}_i v + D_i s  - \tilde h_i\right) - \tilde \lambda_i u_i,  \quad i=1,..., M-1
\label{bbc1}
\end{equation}
\begin{equation}
\frac{dv_j}{dt} =- \lambda_j v_j +
r_j {\bf C}_j u
 - h_j ,
\label{bbc2}
\end{equation}
where $i=1,..., M+1$, $j=1,..., N$.

Equations (\ref{bbc1}), (\ref{bbc2}) can $\epsilon$-realize arbitrary systems
(\ref{qs0}) with the parameter $s$ which can be shown as above (see Section 2),
and this completes the proof.

\section{Robust dynamics}

Our definition of robust dynamics is inspired from similar ideas in viability theory (Aubin et al. 2005).
Let us suppose that the dynamics (the global semiflow $S^t$), generates a number of attractors.
Each attractor ${\cal A}$  has an attraction basin $B({\cal A})$, that is an open set
in the phase space.  Assume that our initial data $\phi$ lie in an attractor,
$\phi \in {\cal A}$, and let us add some noise $\xi$ to the dynamics, representing
the effect of the environment. Trajectories become random, and then it is possible that,
under this noise, the trajectory leaves $B({\cal A})$.

We can now define the following characteristic of stability
under the noise. Let us denote $P(T, B({\cal A}), \phi)$ the probability
that the trajectory $u(t, \phi)$ such that $u(0)=\phi \in {\cal A}$ stays in
$B({\cal A})$ within the time interval $[0, T]$.
\vspace{0.2cm}

{\bf Definition.}
{\em Let us consider a network dynamics
depending on some parameters $\cal P$ and on the noise $\xi$.
We say that the dynamics is robust under the noise $\xi$, if
for each $T$ and $\delta > 0$ there is a choice of the parameters such that
$$
  P(T, B({\cal A}), \phi) < \delta
$$
for each attractor $\cal A$ and $\phi \in {\cal A}$.
}
\vspace{0.2cm}

\subsection{Centralized motif with noise}

For the rest of this section we consider a simplified network, with a single
central node interacting with many satellites. This motif can appear as a subnetwork
in a larger centralized network. In order to study robustness, we consider the
case when the satellites and the center are under the influence of noise.
More general situations, including perturbations of several centers and satellites,
will be studied elsewhere.

The network dynamics is described by the following equations:
\begin{equation}
\frac{\partial u_i}{\partial t} =d_i \Delta u_i -\lambda_i  u_i +  \sigma(b_i v - h_i   + \xi_i(x,t)), \quad i=1,..., N,
\label{Net1}
\end{equation}
\begin{equation}
\frac{\partial v}{\partial t} =d_0 \Delta v -\lambda_0 v +  \sigma(\sum_{i=1}^N a_i u_i - h_0 +\xi_0(x,t)),
\label{Net11}
\end{equation}
The random fields $\xi_i(x,t)$ summarize the effect of various extrinsic noise sources.
These can be random variations of the morphogen, or environment noise, or
genetic variability affecting network interactions.

Intrinsic noise, resulting from stochastic gene expression, could be represented as supplementary
terms outside the sigmoid function. In order to avoid further some tedious
technical difficulties, we postpone the discussion of intrinsic
noise to future work.  Some aspects of the
robustness of patterns with respect to intrinsic noise was studied
numerically by Scott et al. (2010).

The flexibility of such simple networks results from the preceding
section.

We can formulate the following problem: how to choose a
network motif, robust under a given environmental noise, and simultaneously flexible?
The choice can result either from genetic changes (for instance mutations, deletions or duplications of
DNA regions) or from network plasticity (epigenetic changes, such as methylation and chromatin remodeling).

Assume that the considered process is a choice of satellites
   $u_i$ from a large pool of possible regulators.
 We can present this process as a choice
 of $n$ indices ${j_i}, i=1,...,n$ from a larger  set $I_N=\{1,2,...,N \}$ of indices,
where $N \ge n$.
This choice can be done by boolean variables $s_j$ that multiply the coefficients $a_j$: the $j$-th
reagent participates in the network
if $s_j=1$ and does not participate if $s_i=0$. Let us make an important assumption allowing us
to obtain a thermodynamical limit as $N \to \infty$. We assume that
\begin{equation}
 |a_i| < CN^{-1}, \quad N \to \infty.
 \label{aih}
\end{equation}

Now we transform  eqs. (\ref{Net1}),(\ref{Net11}), using the results of the previous subsection.
Let $v=q(x)$ and $u_i=U_i(x)$ be equilibrium solutions of (\ref{Net1}),(\ref{Net11}) where the noises $\xi_i$ are removed.
We suppose that the assumptions of the previous subsection hold.
Let us set
$$
v=q +\tilde v,   \quad u_i=U_i + \tilde u_i,
$$
and $U, u$ denote vectors $(U_1, ..., U_N)$, $(u_1, ..., u_N)$.
Let us set temporarily $\xi_0=0$ ( below we shall show how one can
 stabilize the system state, when $\xi_0 \ne 0$).
  This gives
 \begin{equation}
\frac{\partial \tilde v}{\partial t} =d_0 \Delta \tilde v -\lambda_0  \tilde v + \sigma(
 \rho(U+ \tilde u) - \bar h) - \sigma(\rho(U) - \bar h),
\label{Net3L}
\end{equation}
where we use, for brevity, the notation
$
\rho( u)=\sum_{i=1}^{N} s_i a_i  u_i.
$

The second part of equations takes then the form
\begin{equation}
\frac{\partial \tilde u_i}{\partial t} =  - d_i \Delta \tilde u_i -\lambda_i \tilde u_i +
 \sigma(b_i (q+ \tilde v)+ \xi_i -h_i) - \sigma(b_i q - h_i).
\label{Net4L}
\end{equation}

 To investigate  equations (\ref{Net3L}), (\ref{Net4L}), we use
 a special method justified in a  rigorous way in Appendix. This holds
 under the following assumption:
 \vspace{0.2cm}

 {\bf Assumption 4.4.} { The ``morphogenetic'' noises $\xi_i(x,t)$ are independent on $t$:
  $$
 \xi_i(x,t)= \xi_i(x), \quad i=0,1. ..., N.
 $$
 The functions $\xi_i$ are continuous in $x$ and  a priori bounded
 \begin{equation}
 \sup_{x \in \Omega} |\xi_i(x)| < C_*, \quad i=0,1,..., N.
 \label{uni}
 \end{equation}
 where  a positive constant $C_*$ may be large but it is independent of $N$ for large $N$.
 }
\vspace{0.2cm}

 Notice that Assumption 4.4 guarantees global existence of solutions $\tilde u_i(x,t), \tilde v(x,t)$
 of eqs. (\ref{Net3L}), (\ref{Net4L}) for all $t > 0$.

 Intuitively, one can expect that the term $\tilde v$ in (\ref{Net4L}) in $\sigma$ is small and
 can be, thus, removed. Following this idea,
 let us introduce
 $\eta_i$ as solutions of
 \begin{equation}
\frac{\partial \eta_i}{\partial t} =  - d_i \Delta \eta_i -\lambda_i \eta_i +
 \sigma(b_i q+ \xi_i -h_i) - \sigma(b_i q - h_i).
\label{eta}
\end{equation}
If $\xi_i$ are independent of $t$,
 and sufficiently regular in $x$ then arguments of the previous section show
 that for large $t$
 \begin{equation}
 \eta_i(x,t) \to \bar \eta_i(x),
 \end{equation}
 where $\bar \eta_i$ are solutions of elliptic equations
 \begin{equation}
d_i \Delta \bar \eta_i + \lambda_i \bar \eta_i =G_i(\xi_i(x)), \quad
 G_i(\xi_i(x))=\sigma(b_i q+ \xi_i -h_i) - \sigma(b_i q - h_i)
\label{eta2}
\end{equation}
 under zero Neumann boundary conditions.

 Let us consider equation (\ref{Net3L}) for $v$. Assume  $\rho$ is small. Then we can linearize
 the nonlinear contributions in the right hand side of this equation:
 $$
\sigma(
 \rho(U+ \tilde u) - \bar h) - \sigma(\rho(U) - \bar h)= \sigma'(U - \bar h) \rho(\tilde u) + O(\rho(\tilde u)^2).
$$
We assume that $\tilde u_i$ are close to $\bar \eta_i$. Thus, $\rho(\tilde u) \approx \rho(\bar \eta)$.
Calculations presented in the Appendix show that the  fluctuation influence can be  estimated through the quantity
\begin{equation}
\delta(s,T)= \sup_{t \in [0, T]} {\bf H}(s,t), \quad {\bf H}(s,t)= ||\rho(\bar \eta)||^2.
\label{hamt}
\end{equation}

If $\xi_i$ are independent of $t$, for large $t$ one has
\begin{equation}
{\bf H}(s,t) \to \bar {\bf H}(s)=
||\rho(\bar \eta)||^2.
\label{hamtinf}
\end{equation}

Notice that  $\bar {\bf H}$ can be rewritten in the form
 \begin{equation}
\bar {\bf H}(s)= \sum_{m=1}^{N} \sum_{m'=1}^{N} W_{mm'} (\bar \eta(\cdot)) s_m s_{m'},
\label{spin}
\end{equation}
where $W_{mm}$ are random and
\begin{equation}
   W_{mm'}(\bar \eta(\cdot)) =  a_{m} a_{m'}  \langle \bar \eta_m , \ \bar \eta_{m'}\rangle,
     \label{Wmm}
\end{equation}
here $\langle f, g \rangle$ denotes the inner scalar product in $H$:
$\langle f, g \rangle=\int_{\Omega} fg dx$, where $dx$ is the standard Lebesgue measure.
\vspace{0.2cm}

\subsection{Hard combinatorial problems in network evolution}

We assume that Assumption 4.4 holds.
The minimization of $\bar {\bf H}(s)$ with respect to $s$ should be done under the condition that at least one  satellite is involved, i.e.,
\begin{equation}
R_0(s)=N^{-1}\sum_{i=1}^N s_i  > 0.
\label{restr1}
\end{equation}
The analysis of the minimization problem for this random Hamiltonian is a computationally hard problem
 advanced firstly by methods from statistical physics of spin glasses (see, for example, (Mezard Zecchina 2002)
 for applications to hard combinatorial problems, and (Talagrand 2003) for rigorous justification).
To make the analogy with spin glasses more transparent, we can make change $s_i=2S_i+1$, where spin variables
$S_i$ take values $1$ or $-1$. However, our problem is even more complicated because, besides (\ref{restr1}),
some other restrictions should be taken into account.

In addition to (\ref{restr1}), we must take into account restrictions connected with generation of
several steady states $q_1$, $q_2$, ..., $q_M$, to provide flexibility.
Let us take a small $\epsilon > 0$. By adjusting $s_i$ we would like to obtain a set of equilibria close
to $q_l$.
This gives the following restrictions
\begin{equation}
\sup_{x \in \Omega}\sigma(\sum_{i=1}^N s_i a_i \sigma(b_i q_l -\bar h)- \lambda_0 q_l - h_0) < \epsilon, \quad l=1,2,...,M
\label{restr2}
\end{equation}
or, in a simpler form,
\begin{equation}
\sup_{x \in \Omega}|R_l(s,x) -  B_l(x)| < c\epsilon, \quad l=1,2,...,M
\label{restr3}
\end{equation}
where
$$
R_l=\sum_{i=1}^N M_{li}s_i,
$$
$$
M_{li}= a_i \sigma(b_i q_l -\bar h), \quad B_l =\sigma^{-1} ( \lambda_0 q_l - h_0).
$$
Although $R_l$ are linear in $s_i$ functions, the left hand side of
(\ref{restr3}) is, in general, a nonlinear function of a complicated form.
To overcome this difficulty, we replace the $\sup$ in (\ref{restr3}) by the $L_2$- norm
that gives  quadratic in $s$ functionals:
\begin{equation}
{\bf R}_l(s)=||R_l(s,x) -  B_l(x)||^2 < c\epsilon^2. \quad l=1,2,...,M
\label{restr4}
\end{equation}

We use Lagrange multipliers  $\beta_l$ to take into account conditions (\ref{restr4}). This leads to
the following Lagrange function:
\begin{equation}
{\bf F}(s)=  {\bf H}(s)  + \sum_{l=1}^L \beta_l {\bf R}_l(s)^2.
\label{F}
\end{equation}
Let us remind that the matrix $W_{mm'}$, that determines our hamiltonian ${\bf H}$, is
a random matrix depending on
random fields $\xi_i(x)$ through $\rho(\bar \eta)$.
Let us consider these fields as elements of the Banach space $C^0(\Omega)$ of all bounded continuous
in $x$ vector valued functions.
Let $\mu_{\xi}$ be  a probability measure  defined on the subset of  all such functions satisfying (\ref{uni}).

We also propose that variables $s$ are chosen by a stochastic algorithm. The stochastic algorithm depends
on some set of parameters ${ P}$ that can be adjusted.
Let $\mu_{P}$ be a probability
measure associated with this algorithm (this measure is defined below).

We would like to have a small value of ${\bf F}$ for a ``most part'' of field $\xi$ and $s$ values, with respect to the product measure  $\mu = \mu_{\xi} \times \mu_{P}$.

Finally, the  combinatorial problem can be formulated as follows:
 for a small number $\delta$, find parameters $P$ such that  the probability (computed
 by the measure $\mu$),
\begin{equation}
Prob \{  {\bf F}(\xi(\cdot), s) > \delta \}
\label{minH}
\end{equation}
 is small enough.

\subsection{Mean field solution can be obtained by quadratic optimization}

We show here that the optimization problem is feasible when $N$ is large.
To this end, we define the mean field Lagrange
function $\bar {\bf F}$ that is
 obtained from  ${\bf F}(\xi(\cdot), s)$ by averaging with respect to $\mu$.
In order to estimate the deviations of ${\bf F}$ from $\bar {\bf F}$ we use
 the Chebyshev inequality:
\begin{equation}
P(F,s)=Prob \{  |{\bf F}(\xi(\cdot), s) - \bar {\bf F}(s)| > a \}  \le a^{-2} Var {\bf F},
\label{Che}
\end{equation}
where the probability, the average and the variance should be computed by $\mu$.

The stochastic algorithm for choosing the satellites can be a simple Bernoulli scheme.
Namely, let us consider $s_i$ as mutually independent random variables such that
$$
Prob\{ s_i= 1 \}=p_i.
$$
Thus, the mean field Lagrange function reads
\begin{equation}
\bar {\bf F}(p) = \sum_{i=1}^N  \sum_{j=1}^N \bar W_{ij}p_i p_j +  \sum_{l=1}^L \beta_l {\bf R}_l(p),
\label{hamp}
\end{equation}
where $\bar W_{ij}$ is obtained from $W_{ij}$ by averaging with respect to $\mu_{\xi}$.
Our main idea is as follows.

{\em Step 1: Quadratic programming for the mean field Lagrange function}

First, we minimize $\bar {\bf F}$ with respect to $p_i$. This is a quadratic programming problem
that can be solved in polynomial time.

{\bf QP} {\em  to find a minimum $\bar {\bf F}(p)$ under conditions}
\begin{equation}
0 \le p_i \le 1,
\label{qpr1}
\end{equation}
\begin{equation}
R_0(p)=\sum_{i=1}^N p_i > 0.
\label{qpr2}
\end{equation}

The last condition is trivial and can be omitted.  Therefore, we look for a minimum
of a positively defined quadratic form on the multidimensional box.
The well-known L.Khachiyan {\em ellipsoid algorithm} for this problem runs in $Poly(N)$ time. This proves
such a lemma:
\vspace{0.2cm}

{\bf Lemma 4.5.}  {\em If a solution of the problem {\bf QP} exists, then it can be found in $Poly(N)$ time.}
\vspace{0.2cm}

{\em Step 2: Obtain a small variance of the Lagrange function in the limit N large}

Let us suppose that  $\bar {\bf F} < \delta/2$. Then, using (\ref{Che}) we get

\begin{equation}
Prob \{  {\bf F}(\xi(\cdot), s) > \delta \} <
Prob \{  |{\bf F}(\xi(\cdot), s) - \bar {\bf F}(s)|  > \delta/2 \} \leq 4\delta^{-2} Var {\bf F} .
\end{equation}

Now let us estimate $Var {\bf F}$. We consider $Var {\bf H}$, the rest terms ${\bf R}_l$
can be considered in a similar way. First we estimate variation with respect to $s$
by the measure $\mu_P$. One notices that

$$
Var {\bf  H} =E\sum_{i j i' j'} s_i s_j s_{i'} s_{j'} W_{ij} W_{i'j'} -
E\sum_{i j}  s_i s_j W_{ij} E \sum_{i' j'} s_{i'} s_{j'}  W_{i'j'}.
$$

Notice that if $i \ne i'$ and $j \ne j'$ then
$$
E s_i s_j s_{i'} s_{j'} =
E  s_i s_j  E s_{i'} s_{j'}.
$$
Moreover, $|W_{ij}|=O(N^{-2})$ due to our assumption $(\ref{aih})$
on $a_i$ and Assumption 4.4. Thus we have maximum $N^3$ of non-zero terms in
$DH$, which have the order $O(N^{-4})$.
Thus, the complete variation satisfies
\begin{equation}
Var {\bf H} < C_0 N^{ -1},
\label{disp1}
\end{equation}
where $C_0$ is uniform in $N$ as $N \to \infty$.

Thus, for large $N$ one has $Var {\bf F} \to 0$, thus the probability  (\ref{minH})
is arbitrarily small. This shows that the problem of minimization (\ref{Che}) is feasible in polynomial time $Poly(N)$,
when $N$ is large enough. More
precisely, we have the following
\vspace{0.2cm}

{\bf Proposition 4.6.} {\em If a solution of the problem {\bf QP} exists
and $N$ is large enough, then a solution $s$ satisfying all restrictions  and
minimizing $F$ at level $\delta$ with a probability,
arbitrarily close to $1$, can be found in $Poly(N)$ time.}
\vspace{0.2cm}

{\em Remark}. Above we have studied the case $\xi_0=0$. For smooth $\xi_0(x)$
we can obtain a robustness with respect to $\xi_0$ variations in  a simple way. Namely, for large $N$ one can choose
the constant $C$ in   \eqref{aih}) large enough, then $\sum_{i=1}^N {a_i} u_i -h_0 >> |\xi_0(x)|$.

There arises, however, a natural question: how genetic networks can realize these sophisticated algorithms
which are capable to optimize the network robustness?
A possible answer to this question is that $s_i$ could be themselves involved in a gene network of
the form (\ref{cn1}), (\ref{cn2}).
We showed that gene networks are capable to simulate all structurally stable dynamics.
The fact that this is equivalent to simulating arbitrary Turing machines and thus arbitrary algorithms follows
from results of Koiran and Moore (1999).

\section{Conclusion}

We are concerned with dynamical properties of networks with two types of nodes. The $v$-nodes,
called centers,
are hyperconnected and interact one to another via many $u$-nodes, called satellites.
We show, by rigorous mathematical methods, that this centralized architecture,
widespread in gene networks, allow to realize two fundamental biological strategies:
flexible and robust bow-tie control and Wolpert positional information concepts.

We show how a combination of these strategies leads to the remarkable possibility
to create a ``multicellular organism'', where each ``cell'' can exhibit a complicated time
behaviour, different for different cells.
Centralized network architectures provide the
flexibility important in developmental processes and for adaptive functions.

Contrary to previous works on centralized boolean networks (Aldana 2003), we show that arbitrary
bifurcations between attractors can be controlled by action on satellites, instead
of actions on centers.

To check the robustness of such architectures we have considered a simplified example of a centralized
network with a single center. Such system produces many equilibria, and this dynamical structure can be
protected against large space dependent, random perturbations. We show that in general, designing
an optimal network that is protected against such perturbations boils down to finding the
minimum energy of a spin glass hamiltonian, which is a computationally hard problem. However,
for a large number of satellites, the randomness is filtered and reliable protection against
perturbations results as a solution to a quadratic programming problem, that can be solved in polynomial time.
We expect that similar results hold more generally, for networks with any number of centers.
This suggests an evolutionary bias towards centralized networks where hubs are subjected
to control from many satellites.

These findings can be interpreted in terms of gene networks.
The flexibility control by satellites, and not by transcription
factors (centers) can be a major property
of such networks. It may be easier to act on a
satellite (by silencing or reactivating it), then to perform similar
actions on a center (deletion of a hub proves most of the time to be lethal).
We have proposed miRNAs and CREMs as possible candidates for satellite nodes in
gene networks controlling pattering in development. A few examples of
such centralized motifs are known, such for instance the enhancer system
of the {\em even-skipped} gene of Drosophila (Ludwig et al 2011).
The process of reconstruction of such networks is only at the
beginning (see for instance (Berezikov 2011)).
One could expect that many more examples of centralized
motifs and networks will be found during this process.

{\bf Acknowledgements}. The authors are grateful to John Reinitz, Maria Samsonova and Vitaly Gursky for useful
discussions. We are thankful to M. S. Gelfand and his colleagues
for stimulating discussions in Moscow.

SV was supported by the Russian Foundation for Basic Research (Grant Nos. 10-01-
00627 s and 10-01-00814 a) and the CDRF NIH (Grant No. RR07801) and by a visiting
professorship grant from the University of Montpellier 2.
\vspace{0.2cm}


{\bf Appendix: Proofs and estimates}
\vspace{0.2cm}

{\bf I. The proof of Proposition 2.1}
\vspace{0.2cm}

To outline the proof, let us notice that our system has a typical form, where slow ($v$) and fast ($u$) components are separated:
\begin{equation}
v_t = \kappa F(v, u), \quad u_t=Au + \kappa G(v).
\label{slowfast1}
\end{equation}
Let us present $u$ as $u=U + \tilde u$, where
$U=-\kappa A^{-1} G(v)$ and $\tilde u$ is a new unknown. Let us notice that $U_i$ are solutions
of (\ref{cn6}) under boundary conditions (\ref{Neumann}) and that
$|U_i| < c\kappa$.

By substituting  $u=U + \tilde u$ into (\ref{slowfast1}), we obtain
 \begin{equation}
v_t = \kappa F(v, U +\tilde u), \quad \tilde u_t=A \tilde u + \kappa^2 G_1(v, U+ \tilde u),
\label{slowfast2}
\end{equation}
where
$G_1 =\kappa^{-1} A^{-1} G'(v) v_t=A^{-1} F(v, U +\tilde u)$, the operator $A=diag\{\tilde d_i \Delta - \tilde \lambda_i\}$.
Let us  show that
$G_1(\tilde u, v)$ is a uniformly bounded map in  $\cal H$ for all $u, v$ satisfying a priori estimates
(\ref{est0}). For sufficiently smooth initial data $\phi, \tilde \phi \in C^2$
these estimates and evolution equation (\ref{eveq}) imply
\begin{equation}
||v(t)||_{\alpha} \le C_1,  \quad       t  \ge 0, \ \alpha \in (0,1).
\label{apal}
\end{equation}
The Sobolev embedding gives then
\begin{equation}
||\nabla v(t)||_{L_4(\Omega)} \le c||v(t)||_{\alpha} \le C_2,  \quad       t  \ge 0, \ \alpha \in (1/2,1).
\label{apal}
\end{equation}
To estimate now $G_1=(w_1,...,w_N)^{tr}$, let us notice that $w_i$ satisfy the following equations:
 \begin{equation}
 (\tilde d_i \Delta - \tilde \lambda_i) w_i =g_i(x,v) (d_i\Delta -  \lambda_i) v_i,
\label{estw}
\end{equation}
where $g_i$ are  smooth  functions with uniformly bounded derivatives.
Our goal is, thus, to estimate $||\nabla w||$ through $||\nabla v||_{L_4}$ and
$||v||_{\alpha}$.
Let us multiply (\ref{estw}) through $w_i$ and then integrate  the left hand and the right hand sides
of the obtained equations by parts.
We find
\begin{equation}
 ||\nabla w_i||^2 \le c_1 ||\nabla w_i|| ||\nabla v_i|| + c_2 \langle (\nabla v)^2, |w| \rangle.
\label{estw2}
\end{equation}
To estimate $\langle (\nabla v)^2, |w| \rangle$, we use the Cauchy-Schwartz inequality
$$
|\langle (\nabla v)^2, |w| \rangle| \le c ||\nabla v||_{L_4(\Omega)} ||w||,
$$
 Now we can apply (\ref{apal}) and
 the Cauchy inequality with a parameter $a > 0$ that gives
 \begin{equation}
 ||\nabla w|| \ ||\nabla v||  \le c_1 a ||\nabla w||^2 + C a^{-1} ||\nabla v||^2,
\label{estw3}
\end{equation}
 and if $a > 0$ is small enough ($c_1 a < 1$), we obtain, by (\ref{estw2}) and (\ref{estw3}), the need estimate:
 $$
 ||\nabla w|| < C_3.
 $$
 The second equation in (\ref{slowfast2})
then entails
$$
||\tilde u||_t \le - \beta||\tilde u|| + \kappa^2 \sup ||G_1||,
$$
where $\beta=\min\{\tilde \lambda_i \} > 0$ is independent of $\kappa$.
This gives
$$
||\tilde u(t)|| \le ||\tilde u(0)|| \exp(-\beta t) + C_4 \kappa^2.
$$
In a similar way one can obtain the same estimate for
$||\nabla \tilde u||$.
This completes the proof.
\vspace{0.2cm}

{\bf II. Estimates for network viability via spin hamiltonian}
\vspace{0.2cm}


Assume that for some  $\xi(x)=(\xi_1(x),..., \xi_N(x))$ there holds
\begin{equation}
  {\bf H}(x, \xi(\cdot)) < \delta.
\label{A1}
\end{equation}
Let us obtain   estimates of deviations $\tilde v=v -q$ and $\tilde u_i=u_i - U_i(q)$,
where $v=q(x),\  u=U_i$ define an equilibrium stationary solution for  $\xi_i(x)\equiv 0$.
These estimates hold only due to the special structure of our network: we admit
that $\xi_i$ are not small, nonetheless, the summarized effect of these perturbations is small.
We assume
\begin{equation}
\tilde u_i(x, 0)=0, \quad \tilde v(x, 0)=0.
\label{init0}
\end{equation}
Let us present the  functions $\tilde u_i$
as sums $\tilde u_i=\bar \eta_i + w_i$, where $\bar \eta_i$ are defined by (\ref{eta2}).
For $w_i, \tilde v$ we then obtain
\begin{equation}
\frac{\partial \tilde v}{\partial t} =d_0 \Delta \tilde v - \lambda_0 v  + \sigma(
 \rho(U+ \tilde u(\tau)) - \bar h) - \sigma(\rho(U) - \bar h),
\label{Net30L}
\end{equation}
\begin{equation}
\frac{\partial w_i(t)}{\partial t} =d_i \Delta \tilde v - \lambda_i v +   F_i(\tilde v(\tau), \xi)d\tau,
\label{Net40L}
\end{equation}
where
$$
F_i(\tilde v, \xi)=\sigma(b_i (q+ \tilde v) -h_i +\xi_i) - \sigma(b_i q - h_i +\xi_i).
$$
Let us observe that
\begin{equation}
||\tilde F_i|| < c||\tilde v||.
\label{Fd}
\end{equation}
Condition (\ref{A1}) implies that
\begin{equation}
||\rho( \eta(\cdot))||  < \delta.
\label{ass0}
\end{equation}
Let us introduce $||w||$,
by
$
||w||^2 =\sum_{i=1}^N  ||w_i||^2
$
and $|a|$ by
$
|a|^2=\sum_{i=1}^N  |a_i|^2.
$
Then
$$
||\rho(w)|| \le |a| ||w||.
$$
By (\ref{Net30L}), (\ref{Net40L}),  (\ref{Fd}) and (\ref{ass0})
now one obtains inequalities  for
$||\tilde v||, || w_i||$:
\begin{equation}
\frac{d||\tilde v||^2}{2dt}
\le -\lambda_0 || \tilde v||^2  +  c_3( ||\rho(\bar \eta)|| +  |a|||w||),
\label{group}
\end{equation}
\begin{equation}
\frac{d||w||^2}{2dt}
\le -\bar \lambda || w||^2  +  c_4 ||\tilde v||.
\label{group1}
\end{equation}
where $\min_{i} \lambda_i =\bar \lambda > 0$.
Assume that $\min_{i} \lambda_i > 0$  are large enough.
Moreover, for large $N$ the coefficient $c_3|a| < 1$.
Combining (\ref{group}), (\ref{group1}) one obtains the inequality for
$||Y||^2=||w||^2 + ||\tilde v||^2$:
 \begin{equation}
\frac{d||Y||^2}{2dt}\le -\lambda_0 || Y ||^2  + (\lambda_0 -\bar \lambda) ||w||^2 +
 c_3 \delta ||\tilde v|| +  c_5 ||w|| \ ||\tilde v||.
\label{group2}
\end{equation}
We apply now the Cauchy inequality $xy < ax^2 + a^{-1} y^2$  to the two terms in the right hand side
of this last inequality. This gives
\begin{equation}
\frac{d||Y||^2}{2dt}\le -\lambda_0 || Y ||^2  + (\lambda_0 -\bar \lambda) ||w||^2 + a^{-1} ||w||^2
+ c_6 a ||\tilde v||^2
 + c_7 a^{-1} \delta^2
||\tilde v||.
\label{group3}
\end{equation}
We adjust  an $a$ such that  $c_0 a < \lambda_0/2$. If $\bar \lambda$ is large enough,
 (\ref{group3}) gives then
\begin{equation}
 ||Y(t)|| \le c\delta  + ||Y(0)|| \exp(-\lambda_0 t/2).
\label{est11}
\end{equation}
Then (\ref{est11}) implies  that for large $t$ there holds
$$
\sup_{t > 0}||\tilde v(t)||_{\alpha} \le c_9\delta
$$
with a constant $c_9 > 0$. This gives us the need estimate of $\tilde v$ via the spin hamiltonian.

\end{document}